\documentstyle[aps,prl,preprint]{revtex}
\draft
\tighten

\title{Full-folding optical potential for preequilibrium nucleon\\ 
       scattering at low energies}
\author{M. Avrigeanu$^1$, A.N. Antonov$^2$, H. Lenske$^3$,
        and I. \c Ste\c tcu $^1$}
\address{$^1$Institute for Physics and Nuclear Engineering,
                P.O. Box MG-6, 76900 Bucharest, Romania\\
$^2$Institute of Nuclear Research and Nuclear Energy,
                1784 Sofia, Bulgaria\\
$^3$Institut f\" ur Theoretische Physik, Universit\" at Giessen,
        D-35392 Giessen, Germany}

\begin{document}
\maketitle

\begin{abstract}
The real 
part of the optical potential for the nucleon-nucleus scattering at 
lower energies ($E_i<$100MeV) has been calculated
including nucleonic and mesonic form factors by a double folding 
approach. Realistic density- and energy-dependent
effective $NN$-interactions $DDM3Y$, $BDM3Y$ and $HLM3Y$ based on the
Reid and Paris potentials are used in this respect. The effects of 
the nucleon density distribution and the average relative momentum on 
the folded potential have been analysed. A good agreement with the 
phenomenological potential of Lagrange-Lejeune, as well as with the 
parametrization of Jeukenne-Lejeune-Mahaux for both neutron and proton 
double-folded potentials is obtained. The results indicate that the
strongly simplified model interactions used in preequilibrium reaction
theory neglect important dynamical details of such processes.\\
\end{abstract}

\noindent


The nucleon-nucleus optical model potential (OMP) is an important
tool for understanding and interpreting the mechanism of the
interaction between a nucleon and a nuclear system. The explicit forms
of the $t$-matrices and nuclear matter $g$-matrices have been used
through the folding procedures to obtain more realistic OMP for
practical use in the nuclear reaction calculations \cite{satchler79}.
Brieva and Rook \cite{brieva77} used the local density approximation
(LDA) in order to apply to finite nuclei the nucleon-nucleon ($NN$)
$g$-matrix effective interaction appropriate for a pair of nucleons
interacting in infinite nuclear matter. Jeukenne, Lejeune and Mahaux
(JLM) \cite{jeukenne77} recommended a simpler parametrized form of the
isoscalar real component obtained from Brueckner-Hartree-Fock nuclear
matter calculations, to be used for practical nuclear reaction
calculations. Actually, the use of the free nucleon-nucleon
$t$-matrix as a complex effective interaction has already been applied
successfully to nucleon scattering at higher energies ($E_i>$100MeV),
e.g. \cite{arellano95} and references therein.

Microscopic descriptions have become a standard approach for
the description of reactions leading to discrete final states and giant
resonance, respectively \cite{baker97}. However, a different situation
is encountered for more complex nuclear reactions involving multistep 
processes. In this paper, we consider the inclusion of a microscopic 
OMP into the description of preequilibrium emission processes. Hitherto, 
the complexity of such reactions has been considered to inhibit the use 
of microscopic methods. It is common practice to use strongly simplified 
models, for example the simplest 1 fm range Yukawa interaction has been 
chosen to describe the effective $NN$-interaction leading to 
particle-hole excitations in the frame of the Feshbach-Kerman-Koonin 
(FKK) model \cite{feshbach80}, its strength $V_0$ being the only free 
parameter adjusted to reproduce the experimental data. The large 
discrepancies still present at low incident energies ($E_i<$50 MeV) in 
the systematics of the phenomenological $V_0$ values prove the 
necessity to consider a more realistic $NN$-interaction 
\cite{avrigeanu96} which should be chosen consistently with the
corresponding real part of the optical model potential. One may
consider in this respect the density- and energy-dependent effective
DDM3Y, BDM3Y and HLM3Y interactions based on the $g$-matrix elements 
of the Reid and Paris $NN$-potentials \cite{khoa93,khoa94,hofmann98}, 
which seem to be most successful for the nucleon-nucleus scattering 
within the low-density nuclear matter \cite{khoa93}.

The present work concentrates on a comparison of folding OMPs from 
different interactions and density models. The intention is to derive 
a global model which describes reliably elastic scattering over a 
large range of incident energies and target masses. Once such a model 
is available, it can be applied on safer theoretical grounds to the 
various reaction channels of preequilibrium reactions. Obviously, the 
OMP in these channels cannot be determined directly by experimental 
means. A microscopic model will reduce the still persisting 
uncertainties. Moreover, the consistency between the interaction used 
in elastic scattering and the non-elastic processes is guaranteed.

Usually, effective NN interactions are
parametrized in terms of meson exchange-type propagators. e.g. of 
Yukawa shape. Folding calculations for OMPs and transition potentials 
involve an averaging over the intrinsic momentum distribution of the 
target nucleons. Hence, a region of off-shell momenta is covered. 
Under such conditions the use of ``on-shell'' meson exchange 
propagators might lead to spurious effects because contributions from 
large off-shell momenta are overestimated. These effects are well-known 
from calculations of the free NN $t$ matrix \cite{machleidt89} and in 
Brueckner theory, e.g. \cite{gross92,jong98}. In order to avoid these 
uncertainties one has to introduce vertex and nucleon form factors which 
regularize the large momentum region. Their effect is to exclude 
contributions from small distances where the composite structure of 
mesons and nucleons would become visible.

For a OBE-type interaction the momentum space vertex cut-offs correspond 
to remove in coordinate space the singularities of Yukawa functions at 
the origin. The same effect is obtained by folding an r-space form factor 
to a Yukawa function. In our calculations we follow this prescription
and use a double folding procedure with effective $NN$ interactions and a 
form factor describing the intrinsic structure of the nucleon. We expect 
to take into account typical features of realistic $NN$ potentials 
(e.g. \cite{stoks93,stoks94,machleidt89}) of which the regularized 
Reid93 potential \cite{stoks94} includes explicitly a dipole form factor. 

The main aim of this work is to study the effects of both nuclear and
nucleon density distributions. Equally important is a reliable
treatment of the non-local exchange contributions from anti-symmetrization. 
In order to avoid the complications of a non-local optical potential we use 
the local momentum approximation. Different prescriptions
of the related quantities like the kinetic-energy density $\tau(r)$ and 
various approximations
of the average relative momentum $k_{av}(r)$ \cite{campi78,negele72},
on the corresponding double-folding real OMP at lower incident
energies ($E_i<$100 MeV) are investigated in detail.


The general expressions for the direct and exchange double-folding
(DF) real parts of the optical potential in terms of the
nuclear densities $\rho_1({\bf r}_1)$ and $\rho_2({\bf r}_2)$
of the projectile and the target nucleus, respectively, and the
effective $NN$-interaction
$v({\bf s}$=${\bf R}$ + ${\bf r}_2$ - ${\bf r}_1)$ 
are by assuming proportional proton and neutron densities
\cite{satchler79,khoa93},

\begin{mathletters}\label{eq:1}
\begin{equation} \label{eq:1a}
 U_{0(1)}^D(E,{\bf R})=\delta_{0(1)}\delta_{0(2)}
                 \int d{\bf r}_1 \int d{\bf r}_2\:  \rho_1({\bf r}_1)
             \: \rho_2({\bf r}_2)\: v_{00(01)}^D(\rho,E,{\bf s}),
\end{equation}
\begin{equation} \label{eq:1b}
 U_{0(1)}^{EX}(E,{\bf R})=\delta_{0(1)}\delta_{0(2)}
                     \int d{\bf r}_1 \int d{\bf r}_2\:
                     \rho_1({\bf r}_1,{\bf r}_1+{\bf s})
                     \rho_2({\bf r}_2,{\bf r}_2-{\bf s})\:
                     v_{00(01)}^{EX}(\rho,E,{\bf s})\:
                     \exp \left[{{i{\bf k}({\bf R}){\bf s}}
                              \over {M}}\right],
\end{equation}
\end{mathletters}
where $\delta_{0}$=$1$,
      $\delta_{1}$=$(N_1-Z_1)/A_1$,
      $\delta_{2}$=$(N_2-Z_2)/A_2$,
      $M$=$A_1A_2/(A_1+A_2)$,
      ${\bf k}({\bf R})$ is the incident relative momentum, and
      $\rho_1$ and $\rho_2$ in Eq. (\ref{eq:1b}) are the density
      matrices for the projectile and the target nucleus, respectively.
If the projectile is a nucleon, in the point-like approximation, i.e.
for $\rho(r)$=$\delta$($\bf {r}$), these expressions are reduced to
the single-folded (SF) form. In the frame of the LDA the SF procedure
leads to the nuclear matter approximation (NM) \cite{khoa93}.

In this work we have evaluated Eqs. (\ref{eq:1}) in momentum space
\cite{satchler79,khoa93,khoa94} by means of the detailed procedure
developed by Khoa {\it et al.} \cite{khoa94}, and the following
approximations.
The effective $NN$-interaction has been obtained by using 
the isoscalar and isovector components of the
direct $v^D_{00}$, $v^D_{01}$  and exchange parts $v^{EX}_{00}$,
$v^{EX}_{01}$  of the M3Y interaction based on the results of the
$g$-matrix calculations using either Reid or Paris $NN$-potential
\cite{khoa93}. The energy- and density-dependent M3Y effective
$NN$-interaction has been taken in the form

\begin{equation} \label{eq:2}
 v^{D(EX)}_{00(01)}(\rho,E,r) =
 F(\rho) \:g(E) \:v^{D(EX)}_{00(01)}(r) .
\end{equation}
Three forms of the effective $NN$-interaction density dependence,
namely DDM3Y1 and BDM3Y1 \cite{khoa94}

\begin{mathletters}\label{eq:3}
\begin{equation} \label{eq:3a}
F(\rho)=\left\{ \begin{array}{ll}
C [1+\alpha \exp ( -\beta\rho ) ], & \mbox{for DDM3Y1,} \\
C( 1 - \alpha \rho ), & \mbox{for BDM3Y1,}
\end{array} \right.
\end{equation}

\noindent
and the third one HLM3Y \cite{hofmann98} with different density 
dependence for the isoscalar and isovector components

\begin{equation} \label{eq:3b}
F(\rho)=\left\{ \begin{array} {ll}
f_0(\rho)=s_0 [1 +a_1^0\: (\rho/\rho_0)^{1/3}\:
                 +\: a_2^0 (\rho/\rho_0)^{2/3}\:
                 +\: a_3^0 (\rho/\rho_0)^1] \\
f_{\tau}(\rho)=s_{\tau} [1 \:+\: a_1^{\tau} (\rho/\rho_0)^{1/3}]
\end{array} \right.
\end{equation}
\end{mathletters}

\noindent
where involved in the present work.
The energy-dependent factor is taken to be a linear
function of the incident energy per nucleon $E$ of the form
$g(E)$=1-$\gamma\/E$. The coefficient $\gamma$ has the values 0.002
for the M3Y-Reid interaction, and 0.003 for the M3Y-Paris interaction.

The frozen-density approximation \cite{satchler79,khoa93,khoa94} has
been adopted for the overlap density which enters the explicit form
of the density-dependent factor $F(\rho)$, being taken as the sum of
the densities of the two colliding nuclei at the midpoint of the
intranucleonic separation.
A widely used approximation for the calculation of the knock-on
exchange term of the folded potential is that proposed by Campi
and Bouyssy \cite{campi78}, which preserves the first term of the
expansion given by Negele-Vautherin \cite{negele72} for the realistic
density-matrix expression

\begin{equation} \label{eq:4}
 \rho({\bf R},{\bf R}+{\bf s})=\rho({\bf R}+{{{\bf s}}\over{2}}) \:
       \hat{j}_1\left(k_{av}({\bf R}+{{{\bf s}}\over {2}}) \: s\right) ,
\end{equation}

\noindent
where $\hat{j}_1(x)=3\:(\sin\:x\: - \:x\:\cos\:x)/x^3$, and
$k_{av}$ defines the average relative momentum \cite{campi78}

\begin{equation} \label{eq:5}
k_{av}(r)=\left[{{5}\over{3 \rho(r)}} \left(\tau(r)-
      {{1}\over{4}}\nabla^2\rho(r)\right)\right]^{{1}\over{2}} .
\end{equation}
The last quantity is a function of the density distribution $\rho(r)$,
and the approximation used for the kinetic-energy density $\tau(r)$
for each participant in the interaction.


The Fermi-type nucleon density distribution corresponding to the
parameters $\rho_0$, $R_{1/2}$, and $a$ given by either the
Skyrme-Hartree-Fock systematics \cite{lenske97} or Negele
parametrization \cite{negele70} are compared in Figs. 1(a) and (b).
The Fermi distribution for the target nucleus density, obtained by
using the Negele parametrization \cite{negele70} is shown in Fig.
1(c).

Different forms of the kinetic-energy density $\tau(r)$ have been
adopted for the light projectiles and for the target nuclei. Thus,
the modified Thomas-Fermi (MTF) approximation \cite{krivine79}
has been considered for light projectiles

\begin{equation} \label{eq:6}
\tau(\rho)=\alpha  \rho(r)^{{5}\over{3}} +
           \beta {{|(\nabla \rho)|^2}\over{\rho}} .
\end{equation}
The values of the parameters $\alpha$ and $\beta$ depend on the number
of the constituent nucleons of the projectiles. Following the analysis
of Krivine and Treiner \cite{krivine79}, we used for the nucleon as a
projectile $\alpha=0$ and $\beta=1/4$. For heavier
nuclei and also within the MTF frame we have
$\alpha=3(3\pi^2)^{2/3}/5$. Consequently, the following expression
have been obtained for $k_{av}(r)$ in the case of light projectiles
(here nucleons)

\begin{mathletters}\label{eq:7}
\begin{equation} \label{eq:7a}
 k_{av}^{MTF}(r)=\left[{{5 |\nabla\rho(r)|^2}\over{12 \rho^2(r)}}
  -{{5 \nabla^2\rho(r)}\over{12 \rho(r)}}\right]^{{1}\over{2}} ,
\end{equation}
while for heavier nuclei we have

\begin{equation} \label{eq:7b}
 k_{av}^{MTF}(r)=\left[ k_F(r)^2
               +{{5 |\nabla\rho(r)|^2}\over{12 \rho^2(r)}}
  -{{5 \nabla^2\rho(r)}\over{12 \rho(r)}}\right]^{{1}\over{2}} ,
\end{equation}
\end{mathletters}
where $k_F(r)=[3\pi^2\rho(r)/2]^{1/3}$ is the local Fermi momentum.

On the other hand, we have also used the extended Thomas-Fermi (ETF)
approximation \cite{campi80}

\begin{equation} \label{eq:8}
\tau(\rho)=\alpha  \rho(r)^{{5}\over{3}} +
           \beta {{|(\nabla \rho)|^2}\over{\rho}} +
           \gamma \nabla^2 \rho ,
\end{equation}
where the widely-used values of the parameters $\alpha$, $\beta$,
and $\gamma$ \cite{campi80}

\begin{equation} \label{eq:9}
  \alpha={{3}\over{5}} \left( 3 \pi^2 \right)^{{2}\over{3}} , \: \:
  \beta={{1}\over{36}} , \: \: \gamma={{1}\over{3}}
\end{equation}
lead to the following expression for the average relative momentum

\begin{equation} \label{eq:10}
k_{av}^{ETF}(r)=\left[ k_F(r)^2
               +{{5 |\nabla\rho(r)|^2}\over{108 \rho^2(r)}}
  +{{5 \nabla^2\rho(r)}\over{36 \rho(r)}}\right]^{{1}\over{2}} .
\end{equation}
The MTF kinetic-energy densities for neutrons and protons are shown in
Figs. 1(d) and (e), respectively.
The kinetic-energy densities derived in the ETF and MTF approximations 
are in close agreement for $^{93}$Nb, as seen in
Fig. 1(f). These results prove the consistency of the parametrizations
involved. The related $k_{av}(r)$ curves are shown in Figs. 1(g),(h)
and (i) for nucleons and $^{93}$Nb, respectively. The increased
nucleon $k_{av}$-values at smaller radii follow the behaviour of the
$\nabla^2\rho(r)$ for a Fermi distribution, whose importance in the
nucleon case results from Eq. (\ref{eq:7a}).

Furthermore, the local incident momenta $k_i(r)$ calculated for 20
MeV neutrons and protons on $^{93}$Nb \cite{avrigeanu96} by using the 
phenomenological OMP of Lagrange-Lejeune \cite{lagrange82} as well as 
the actual DF potential are also shown in Figs. 1(g) and (h). The
minor differences in both  
shape and magnitude support the use of the 
Negele-Vautherin approximation \cite{negele72} as Eq. (\ref{eq:4}) 
and the MTF approximation for the calculation of $k_{av}(r)$.
The $k_{av}^{MTF}(r)$ curve for the target nucleus $^{93}$Nb
decreases at larger $r$, however slower than $k_F(r)$.


The real OMPs for incident neutrons and protons on $^{93}$Nb,
calculated by means of (i) the single-folding procedure within the
LDA, (ii) the complete SF procedure, and (iii) the full DF, are
compared with both the phenomenological OMP \cite{lagrange82} and the
JLM parametrization \cite{jeukenne77} in Fig. 2. The DDM3Y1-Paris
effective $NN$-interaction, the Fermi density distribution
\cite{lenske97} and the MTF approximation for $k_{av}(r)$ have been
used in this respect. Moreover, the corresponding volume integrals
$J_V$ and $rms$ radii $\langle r^2\rangle^{1/2}$ are given
in Table 1. The real part of the JLM potential has been chosen for
the present analysis since the parameters of the DDM3Y1 interaction
\cite{khoa93,khoa94} was obtained by fitting of the JLM results for
the nucleon scattering \cite{jeukenne77,khoa93,khoa94}. On the other
hand, the well-known OMP parameter set of Lagrange-Lejeune
\cite{lagrange82} has also been computed for both kinds of projectiles
on $^{93}$Nb, close to the JLM analysis.

First, Fig. 2 shows that more diffuse potentials with larger
$\langle r^2\rangle^{1/2}$ values are obtained for neutrons as well
as for protons when the folding procedure increases in accuracy going
from NM to DF cases. Actually, the comparison in Figs. 2 (a),(b) and
(d),(e) of the real OMP obtained by using the NM and SF methods
evidences the effects of the target nuclear density and the
corresponding average relative momentum. Next, the comparison with
Figs. 2(c) and (f) shows even stronger effect of the nucleon density
distribution and the average relative momentum. On the other hand, the
different choices of the target $k_{av}$ (MTF or ETF) have practically
no effect on the single-folded real OMP, the corresponding curves in
Figs. 2(b) and (e) being overlapped.

Second, the comparison between the real parts of the OMPs shown in 
Fig. 2 to the phenomenological values \cite{lagrange82} and the JLM 
parametrization \cite{jeukenne77} is completed by considering the 
energy dependence of the corresponding volume integrals. From Fig.3 
it is seen that the phenomenological and both calculated $J_V$ values 
from the SF and the DF approach agree rather well. In view of the 
increasing agreement for the depths (Fig. 2) and the 
$\langle r^2\rangle^{1/2}$ values (Table I) in going from the SF to 
DF procedures these comparisons can be considered an important test 
of the present calculations. A comment concerning the proton case is 
necessary. There, the comparison to the JLM parametrization is less 
significant, because the latter accounts only for the isoscalar 
component of the full real OMP. 

Third, we have considered in the present work three different
density-dependent versions DDM3Y1, BDM3Y1 and  HLM3Y of the M3Y-Paris 
and M3Y-Reid $NN$-interactions. The double-folded real OMPs 
corresponding to the former interaction for DDM3Y1 and BDM3Y1 are 
shown in Fig. 4 for the neutron as well as the proton cases. It 
results that the two density dependences lead to almost identical 
folded real OMPs provided that a Fermi-distribution of the nucleon 
density is used. Also, the calculated real part 
of the OMP using HLM3Y interaction \cite{hofmann98} lead to comparable
results with those obtained by DDM3Y1 and BDM3Y1. The radial 
dependence of the real OMP corresponding to the Paris-HLM3Y effective
$NN$-interaction, however evidences enhanced values in the surface 
region as can be seen from the $rms$ radii values in Table 1. On the 
other hand, an obvious difference exists between the DF real potentials 
corresponding to the simple M3Y-interaction and the density-dependent 
versions, respectively.

The density dependence of the effective $NN$-interaction improves the 
agreement of the folded potentials with the phenomenological ones by 
reducing the DF potential values at small values of $r$ where the 
highest overlap of the interacting projectile and target-nucleus 
densities takes place. However, the simple M3Y folded potentials are 
less diffuse in the nuclear surface region, e.g. they have smaller 
$\langle r^2\rangle^{1/2}$ values than the real OMPs obtained by means 
of the energy-dependent $NN$-interactions (see Table 1).

The importance of the realistic effective $NN$-interactions is pointed
out by showing also in Fig. 4 the DF real potentials calculated by
means of the usual 1 fm range Yukawa interaction with the $V_0$ 
strength values obtained by Watanabe {\it et al.} for these systems 
\cite{watanabe95}, i.e., $V_0$=40 MeV for $n$+$^{93}Nb$ and 46.8 MeV 
for $p$+$^{93}Nb$. On the other hand, there are also shown the results
obtained by using the $equivalent$ 1 fm range Yukawa interaction with 
$V_0$ strengths corresponding to the volume integral values provided 
by the Paris-M3Y interaction. Thus, the improvement in comparison with 
the 1 fm range Yukawa interaction obtained just using the simpler M3Y 
interaction is obvious. Rather similar effect has the use of the 
$equivalent$ $V_0$ values, corresponding to the Paris-M3Y volume 
integrals. Obviously, the supplementary density dependence increases 
the agreement with the phenomenological optical potentials [see Figs. 
2(c)(f)] considerably.

Finally, one comment may concern the effect of the $NN$-interaction
type, i.e. of the Paris and Reid potentials. 
As seen from Table 1, the $J_V$ and $\langle r^2\rangle^{1/2}$
values for
the corresponding neutron and proton folded potentials are close to 
each other, independent of using the NM, SF, and DF. Thus one may 
conclude that both $NN$ interactions lead to reliable results in 
folding calculations.

In conclusion, this analysis has shown that the double folding
procedure involving the nucleon-density Fermi distribution and average
relative momentum leads to real OMPs at lower energies in good
agreement with the phenomenological OMPs as well as with the JLM
parametrization for both neutron- and proton-nucleus scattering. The
analysis of the average effective $NN$-interaction strength by using 
the actual folded real OMP and following \cite{avrigeanu96} is in 
progress. Since this quantity is still considered as the only free 
parameter of the quantum-statistical studies of the multistep 
reactions, the corresponding search for the realistic effective 
$NN$-interaction at low energies is thus of further interest.

The comparison of the full folding calculations to the results obtained with the
strongly simplified single Yukawa description is very
instructive showing clearly the restrictions of a "simple"
parameterization. Apparently, the depth of the potential is overestimated
because of the lack of density dependence, as one should expect.
For inelastic transitions at a fixed incident energy one might compensate
that shortcoming by simply rescaling the V$_0$ parameter such that the
"local" interaction strength in the reaction region is accounted for in
the average. However, the
penetration and therefore the interaction region changes with incident
energy and Q-value. Very likely, the observed energy dependence of V$_0$ is
actually produced to a large extent by the underlying
density dependence which was taken into account explicitly in this work.

A clearer account of that effect could be obtained from elastic and
inelastic scattering reaction calculations (OMP and DWBA) at different incident
energies where one tries to reproduce the  cross sections derived with density
dependent interactions by OMPs and form factors from a single Yukawa interaction.

\newpage

Table 1. Volume integrals (in MeV$\cdot$fm$^3$) and radii
$\langle r^2\rangle^{1/2}$ (in fm) for the real part of folded OMP
calculated by means of the nuclear matter (NM),
single-folding (SF) and double-folding (DF) procedures,
phenomenological OMP \cite{lagrange82} and JLM parametrization
\cite{jeukenne77}.\\
\begin{tabular}{lcccccccc} \hline \hline
Interaction & \multicolumn{2}{c}{NM} & \multicolumn{2}{c}{SF} &
   \multicolumn{2}{c}{DF} & \multicolumn{2}{c}{Phenomenological}\\
        \cline{2-9}
&$J_V$ & $\langle r^2\rangle^{1/2}$ & $J_V$ & $\langle r^2\rangle^{1/2}$ &
 $J_V$ & $\langle r^2\rangle^{1/2}$ & $J_V$ & $\langle r^2\rangle^{1/2}$ \\
\hline
 & \multicolumn{8}{c}{neutrons} \\
JLM         & 404.5 & 4.652\\
            &       &       &       &       & & & 398.6 & 4.930\\
Paris-M3Y   & 383.0 & 4.399 & 402.0 & 4.692 & 398.7 & 5.057\\
Paris-HLM3Y & 448.3 & 4.812 & 470.3 & 5.072 & 422.4 & 5.270\\
Paris-HLM3Y$^*$  & 452.9 & 4.857\\
Paris-BDM3Y1 & 385.7 & 4.527 & 404.9 & 4.808 & 384.9 & 5.168\\
Paris-DDM3Y1 & 395.7 & 4.567 & 415.3 & 4.846 & 392.2 & 5.194\\
Reid-DDM3Y1  & 360.2 & 4.568 & 383.7 & 4.854 & 367.0 & 5.206\\
\hline
 & \multicolumn{8}{c}{protons} \\
            &       &       &       &       & & & 470.1 & 4.930\\
Paris-M3Y   & 429.3 & 4.413 & 452.7 & 4.692 & 451.4 & 5.075\\
Paris-HLM3Y & 501.1 & 4.793 & 528.2 & 5.046 & 483.0 & 5.268\\
Paris-HLM3Y$^*$    & 507.7 & 4.839 \\
Paris-BDM3Y1 & 433.1 & 4.539 & 456.9 & 4.808 & 438.0 & 5.183\\
Paris-DDM3Y1 & 444.2 & 4.579 & 468.6 & 4.846 & 446.3 & 5.209\\
Reid -DDM3Y1 & 422.0 & 4.584 & 453.0 & 4.865 & 436.4 & 5.234\\
\hline \hline
\end{tabular}


\bigskip

$^*$The Hartree-Fock nuclear density for $^{93}$Nb has been used in 
this case \cite{lenske97}.

\newpage
\section*{Figure Captions}
\begin{itemize} \itemsep 0pt \topsep 0pt \parskip 0pt
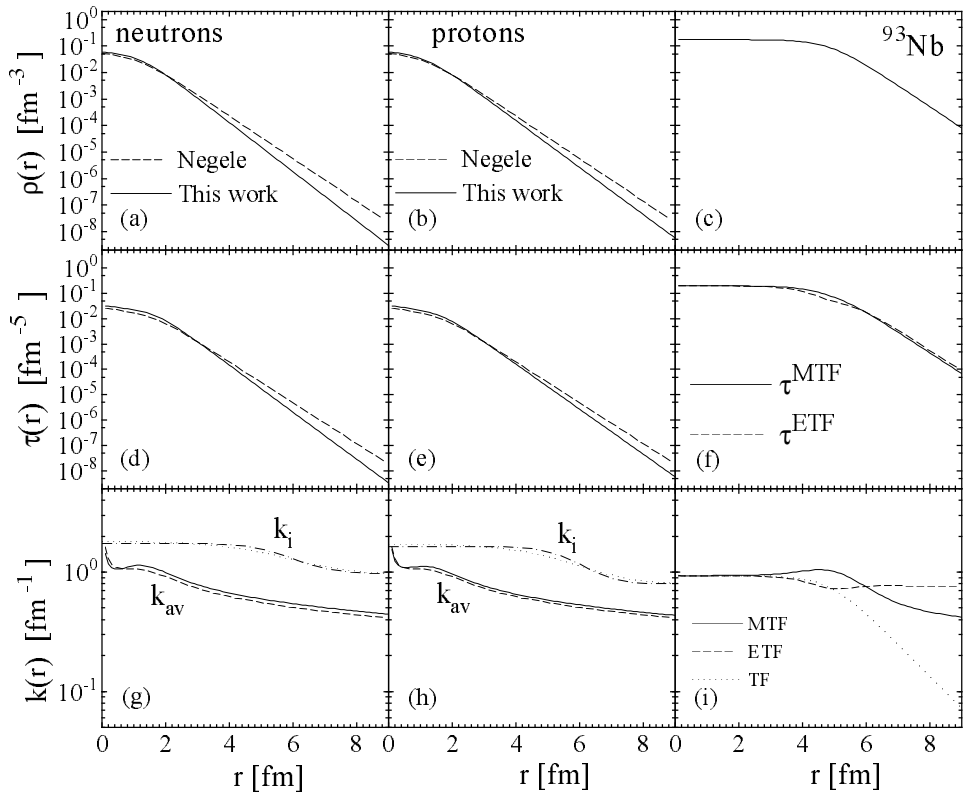
\item [FIG. 1.] Nuclear density distributions, kinetic-energy
densities $\tau(r)$ (MTF, solid curves) and (ETF, dotted curve), the
local incident momenta $k_i$ corresponding to the 20 MeV incident
energy, and MTF average relative momenta $k_{av}(r)$, for (a,d,g)
neutron, (b,e,h) proton, and (c,f,i) the target nucleus $^{93}$Nb,
respectively. The Fermi-distribution as given by Lenske
\cite{lenske97} (solid curves) or Negele parametrization
\cite{negele70} (dotted curves) are used for nucleon density
distribution, and only the latter for the target nucleus. The
$k_i$ values correspond to the phenomenological OMP of
Lagrange-Lejeune (LL) \cite{lagrange82} (dashed-dotted curves) and to
the DF potential (dashed curves), while the MTF average relative
momentum (solid curve), the ETF $k_{av}$ (solid curve), and the local
Fermi momentm (dotted curves) are shown for the taget nucleus.

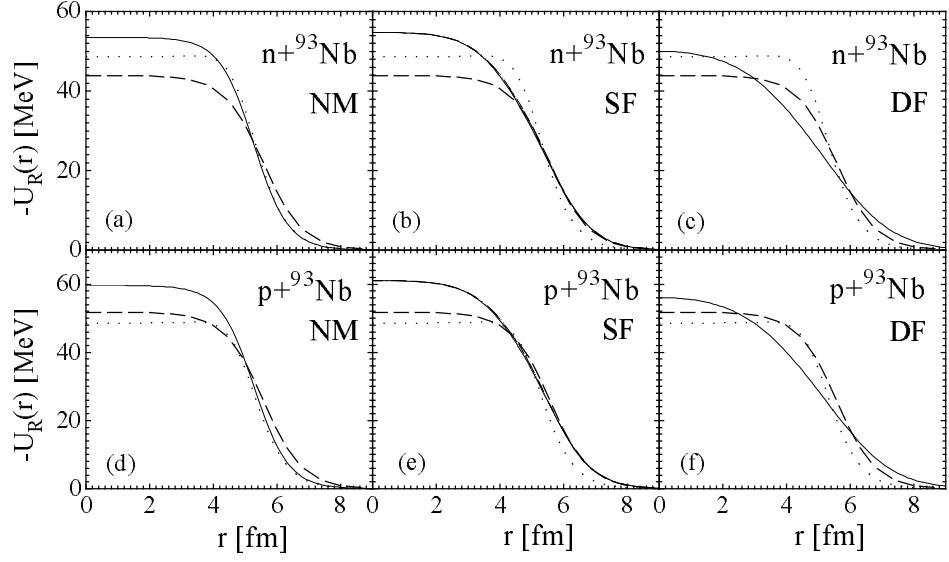
\item [FIG. 2.] The radial dependence of the real OMP calculated with
the Paris-DDM3Y1 effective $NN$-interaction in the MTF approximation,
by using the nuclear matter approximation (NM), single-folding (SF)
and double-folding (DF) procedures (solid curves), compared with the
phenomenological OMP \cite{lagrange82} (dashed curves) and JLM
parametrization \cite{jeukenne77} (dotted curves), for (a,b,c)
neutrons and (d,e,f) protons of 20 MeV incident energy on the target
nucleus $^{93}$Nb.

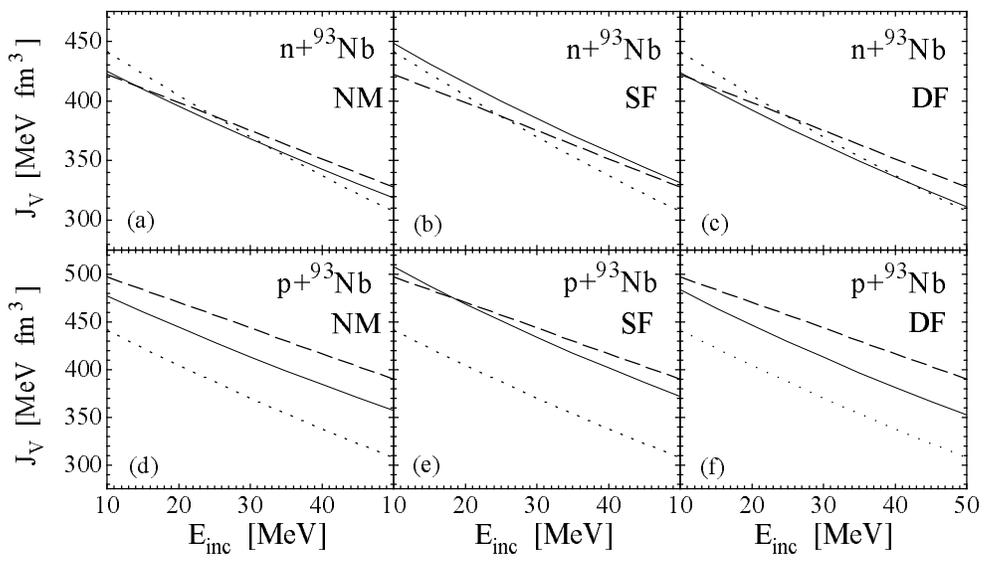
\item [FIG. 3.] The incident-energy dependence of the volume integrals
of real OMP calculated with the Paris-DDM3Y1 effective $NN$-interaction
in the MTF approximation, by using the nuclear matter approximation
(NM), single-folding (SF) and double-folding (DF) procedures (solid
curves), as well as with the phenomenological OMP \cite{lagrange82}
(dashed curves) and JLM parametrization \cite{jeukenne77} (dotted
curves), for (a,b,c) neutrons and (d,e,f) protons incident on the
target nucleus $^{93}$Nb.

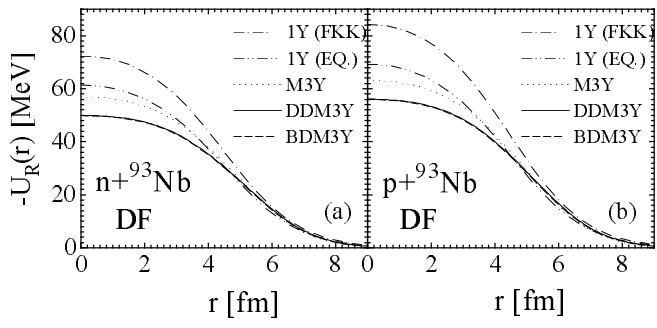
\item [FIG. 4.] The same as Fig. 2, for (a) neutrons and (b) protons,
and corresponding also to the effective $NN$ interactions M3Y (dotted 
curves), BDM3Y1 (dashed curves), 1 fm range Yukawa with the $V_0$ 
values either obtained through FKK analysis \cite{watanabe95} 
(dotted-dashed curves) or equivalent (double dotted-dashed curves) to 
the M3Y interaction (see text).

\end{itemize}
\end{document}